\documentclass[journal]{IEEEtran}

\usepackage{booktabs} 
\usepackage{array}
\usepackage{adjustbox}
\usepackage{tikz}
\usetikzlibrary{positioning}
\usepackage[
  separate-uncertainty = true,
  multi-part-units = repeat
]{siunitx}

\usepackage{amsmath}
\usepackage{algorithm}
\usepackage[noend]{algpseudocode}
\usepackage{amsmath}

\begin{document}
\title{Covfefe: A Computer Vision Approach For Estimating Force Exertion}

\author{Vaneet Aggarwal, Hamed Asadi, Mayank Gupta, Jae Joong Lee, and Denny Yu\thanks{The authors are with Purdue University, West Lafayette, IN USA 47906, email:\{vaneet,hasadi,gupta369,lee2161,dennyyu\}@purdue.edu.} }

\maketitle
\begin{abstract}
Cumulative exposure to repetitive and forceful activities may lead to musculoskeletal injuries which not only reduce workers' efficiency and productivity, but also affect their quality of life. Thus, widely accessible techniques for reliable detection of unsafe muscle force exertion levels for human activity is necessary for their well-being. However, measurement of force exertion levels is challenging and the existing techniques pose a great challenge as they are either intrusive, interfere with human-machine interface, and/or subjective in the nature, thus are not scalable for all workers. In this work, we use face videos and the photoplethysmography (PPG) signals to classify force exertion levels of 0\%, 50\%, and 100\% (representing rest, moderate effort, and high effort), thus providing a non-intrusive and scalable approach. Efficient feature extraction approaches have been investigated, including standard deviation of the movement of different landmarks of the face, distances between peaks and troughs in the PPG signals. We note that the PPG signals can be obtained from the face videos, thus giving an efficient classification algorithm for the force exertion levels using face videos. Based on the data collected from 20 subjects, features extracted from the face videos give 90\% accuracy in classification among the 100\% and the combination of 0\% and 50\% datasets. Further combining the PPG signals provide 81.7\% accuracy. The approach is also shown to be robust to the correctly identify force level when the person is talking, even though such datasets are not included in the training.

\end{abstract}
\section{Introduction}
In the United States, 155 million people work full-time as part of their daily lives. Although all employers are ethically required to provide a safe and healthy workplace for their employees, people are still getting hurt daily. Workplaces injuries like musculoskeletal disorders are preventable, and workplace risk factors are known. However, monitoring these factors reliably and in a scalable way is a key challenge. We propose a computer vision framework for preventing musculoskeletal injuries. The following sections further detail the background motivation, related work, and our contributions.

\subsection{Motivation}
\subsubsection{Work-related MSDs}
  Musculoskeletal disorders (MSDs), such as sprains or strains resulting from overexertion, accounts for 349,050 cases for all workers \cite{U.S.BureauofLaborStatistics2016} annually. This means that 33 workers in every 10,000 suffer an injury severe enough that they must take time away from work. \cite{U.S.BureauofLaborStatistics2016} Although overall percentage of the workforce getting hurt is small, these injuries are preventable. Furthermore, they not only impact individual worker's health and quality of life\cite{wells2009have}, they also result in significant cost employees and society (e.g., workers compensation, medical care, loss productivity, training temporary workers). The annual cost of the injuries in the United States are nearly \$60 billion in direct workers compensation costs \cite{LibertyMutualResearchInstituteforSafety2017}. Due to high direct and indirect cost of MSDs,  there is a strong motivation for all stakeholders (e.g., employers, workers, and researchers) to identify factors that lead to MSDs and actively monitor and eliminate worker exposure to these factors.

\subsubsection{Force Exertion as a predictor of MSDs}
High force exertion levels are reported as the most common contributing factors with sufficient evidence to suggest a causal relationship for work-related musculoskeletal disorders (MSDs) \cite{Buckle2002, Keyserling2000, Schneider2001, Hauret2010, Koppelaar2005}. A comprehensive report by the National Institute for Occupational Safety and Health (NIOSH) lists high/sustained force, repetitive movements, and poor biomechanical postures are contributors to MSDs, with conclusion that evidence exists linking force to musculoskeletal injuries \cite{NationalInstituteforOccupationalSafetyandHealthNIOSH1997}.  

Several key physiological and biomechanical mechanisms are proposed for how force exertions  lead to injuries. For instance, chronic low back pain can be a result of tears in the soft tissues \cite{schwarzer1995prevalence}. For instance, high and/or frequent force exertions initiates lumbar disc damage and degeneration \cite{adams2000mechanical}. In addition to high force exertions, prolonged/sustained force exertions could also lead to work-related MSDs. For example, prolonged force exertions could lead to wrist injuries where frequent force exertions by the hand (e.g., pinching and griping) lead to and exacerbate inflammation of the carpal tunnel cumulative tissue stress can eventually lead to injuries \cite{fung2007study}. 

Although repetition, postures, and vibration are contributors to injuries, \textbf{force} is one of the hardest to measure because it is difficult to observe and depend on individual's effort. For example, changes in expressions are subtle unless high forces and strong efforts are needed. Many methods are currently available to measure the force exertion levels (detailed in Related Work). However, each method vary in reliability and feasibility as they are either 1) intrusive (e.g., disrupts the worker while they are performing their job), 2) interfere with human machine interface (e.g., need to install force gauges on tool-handles and machine controls), 3)  subjective, and most importantly 4)  not widely scalable across all workers, jobs, and workplaces as trained ergonomics and safety professionals are needed to implement these methods.

\subsubsection{Why Computer Vision approach is needed?}
 This paper proposes objective and automated predictions of force levels which has minimum distractions on workers and could be used in wide variety of workplaces by using the videos of the person to predict the level and frequency of force exertions. Innovations in computer vision techniques can address many of the deficiencies in the current approaches. This paper proposes a new objective approach, which can be widely accessible and is not intrusive to workers. 
 




\subsection{Related Work}
Here, we present the existing literature and talk about the different types of force exertions that various researchers have used in the work. To demonstrate this approach, we focus on experiments on force exertions using a grip posture. There are several existing methodologies to estimate the exertion level. The overview of different methods is given here along with related work of various researchers using computer vision in estimating individuals health attributes and activity.

\subsubsection{Types of Force Exertions:} 
Lifting, push/pull, pinch grip, and power grip are common methods for exerting force in workplaces and are especially common in manual handling work tasks \cite{bao2006quantifying}. For example, in manufacturing and manual handling, workers perform tasks throughout the workday using a combination of the aforementioned exertion types. In this study the power gripping force exertion is selected as input force exertion level due to the following reasons. First, hand grip strength is important in designing hand tools\cite{taha2005grip}. The second reason is the engagement of limited muscle in grip force exertion. In comparison to lifting, and pull/push, lower muscle mass is in involved thus potentially places workers performing exertions at the limits of their capabilities \cite{roberts2011review}. Even with lifting, pulling, or pushing, the recommended hand-object coupling is a grip posture. The third reason is that grip exertions are upper-extremity movements that again have lower strength capability than whole-body force exertions. Moreover, focusing on upper-extremity actions limits lower body movements and thus allow for better experimental control to demonstrate our approach.
 
\subsubsection{Assessment of Force Exertion Levels:} 
Various methods have been used by different researchers to measure, estimate workers hand force exertions. For example, the physical exertion level is commonly rated using visual scales \cite{borg1990psychophysical}. Other techniques include estimating hand forces through context, i.e., using the object weights or checking the carrying loads by observing and interviewing the workers \cite{stetson1993median}, measured with a force gauge or mimicked on a hand dynamometer by workers \cite{casey2002getting, bohannon2006reference}, observed by ergonomists, or measured by electromyography on the forearm muscles \cite{keir2005development,sidek2012mapping}.

The conventional method to measure hand grip force  uses hand grip dynamometers \cite{casey2002getting, bohannon2006reference}. Most hand dynamometers operate using strain gauge and directly measures hand grip strength. In workplaces, the workers may be asked to replicate the tasks forces on grip dynamometers. Although these devices provides actual force measurements, their usability is limited due to their availability to workers.

In observational methods, the force levels is observed and estimated by trained ergonomists. The ergonomists are trained to recognize these subtle cues (twist in body, strain in face, perspiration). These signs will become more clear in high force assessing requirements. This method could be also performed on recorded videos of the workers. This method is subjective and based to the estimations \cite{fan2014association}.

Electromyogram (EMG) is a signal that can be measured from the skin surface. Various studies have used the EMG sensors to measure the muscle activation and hand grip strength. The EMG signals measures the activation of forearm muscles. The recorded signals can be filtered and normalized with the maximum activity to represent the hands grip forces \cite{keir2005development,sidek2012mapping, spielholz2001comparison}. This method requires the EMG sensors which are not widely available, cannot be used in workplaces due to time constrains, and are intrusive to workers. 


The proposed computer vision method identified and implemented both sub-surface (PPG, representine blood flow, etc.) and surface (facial expressions) variables to classify the force exertion levels. 


\subsubsection{Use of Computer Vision for Health Attributes:} 
Advances in computer vision and machine learning have the potential to address limitations of current ergonomics state-of-the-art methods for collecting force exertion data in workplace exposure assessments. Automated video exposure assessment has been used in previous studies to automatically quantify repetitive hand activity with the use of digital video processing \cite{Chen2013}. Using video recordings to measure the ergonomic risk factor of repetitive motions, the investigators developed algorithms that tracked the hand and calculated kinematic variables of tasks such as frequency and speed \cite{Greene2017}. The authors of  \cite{Akkas2016} demonstrated that marker-less video tracking algorithm can be used to measure duty cycle and hand activity levels in repetitive manual tasks. The computer vision-based motion capture has been used previously to track and build on-site biomechanical model of the body and minimize work related ergonomic risk factors on construction sites \cite{liu2016tracking, seo2014motion, starbuck2014stereo}. The computer vision approach provides a promising tool for quantifying ergonomic risks from repetitive movements and potentially non-neutral postures; however, force exertion levels are another key ergonomic risk factor that computer vision techniques have not been developed to detect force exertion levels that may associate with injury risks.

Over the last decade, there had been great interest in utilizing the videos of human and estimate the health related parameters like heart rate, breathing rate etc. It is shown that contact methods such as using pulse oximeter can be easily replaced by non contact methods such as human videos \cite{kumar2015distanceppg, poh2011advancements, humphreys2007noncontact, verkruysse2008remote}. All these works utilizes human facial videos and extract relevant features from the human face to estimate health related parameters. The features used are the pixel intensity values from each frame of the video. These features are combined together to generate PPG signal using signal processing techniques and hence estimating health parameter.

\subsubsection{Use of PPG signal for health monitoring:}
Photoplethysmogram (PPG) is an optical technique for the volumetric measurement of the organ. It generates a pulsating wave based on the changes in volume of the blood flowing inside the arteries.   
Recently, there has been growing interest of the researchers in exploring PPG signal. Till date, the PPG signal has been used to extract information such as oxygen saturation level, blood pressure, respiration rate,  heart rate, and heart rate variability. It is also a promising technique that is used in early screening of various atherosclerotic pathologies \cite{elgendi2012analysis}. The amplitude of the PPG has been used as an indicator to vascular distensibility \cite{dorlas1985photo}. This information is used by anesthesiologists to judge subjectively whether a patient is sufficiently anesthetized for surgery. PPG waveform can be a useful tool for detecting and diagnosing cardiac arrhythmias as well. The researchers have also analysed first and second derivate of PPG signal. The first derivative of the PPG can also be used to calculate the augmentation index which is a measure of arterial stiffness \cite{elgendi2012analysis}. The measure of arterial stiffness can be further be related to vascular aging \cite{shirwany2010arterial}. There have been numerous applications of PPG signal and researchers are still exploring the potential of this signal.

\subsection{Contributions}

This work provides first ever approach for prediction the force exertion level using techniques of computer vision and machine learning, to the best of our knowledge. We propose an algorithm that can classify between three different levels of force exertion, i.e., 0\%, 50\%, \& 100\%. Our methodology provides an overall accuracy of over 80\% in correctly classifying these three levels. This algorithm uses efficient feature extraction methods from the facial videos and the PPG signals to perform the classification. The proposed method first uses a classification based on features extracted from the video data to classify between two levels, 100\% and $\le 50\%$. The features from the PPG signals are then used to differentiate between 0\% and 50\%. Since the PPG signals can be obtained from the face videos \cite{kumar2015distanceppg}, this approach is a computer vision approach with the face video as an input and an estimate of force exertion level as the output. Since we are not aware of an existing data set containing face videos and the force exertion levels, we design our own experiment to collect the relevant data for training our model. The data is collected using 20 subjects in total, where each subject was asked to perform different levels of force exertion activities and their videos \& PPG data was recorded. Relevant set of features are extracted from this data and two neural networks are trained to achieve an overall accuracy of 81.7\%.

The second key contribution in our work is the extraction of average movement of facial landmarks from the video data. Such features has never been used to train a machine learning algorithm for predicting force level in the existing literature, to the best of our knowledge. The feature extraction is essential since the large number of features in the face video may cause the classification algorithm to overfit to training samples and generalize poorly to new samples. We demonstrate that the extracted features are the key features in training the model to provide the classification  between high (100\%) \& low (0\% \& 50\%) force exertion level.  We use Deepface algorithm \cite{taigman2014deepface} to process the collected videos. The different frames of the videos are aligned and human face is detected in each frame using Deepface.  The spatial location of  facial landmarks (128 points on the face) are tracked in the entire video and the average movement of each landmark is calculated relative to its position in the first frame of the video. The detailed explanation on such feature extraction is given in Section \ref{sec:landmarks}.

The third contribution in the proposed paper includes efficient feature extraction from  photoplethysmogram (PPG) signals. PPG signal is collected for each subject at every force exertion level during our experiments. We utilize this signal to extract various kind of features as detailed in Section \ref{sec:ppg}. This approach of feature extraction is novel, and has not been studied in prior works to the best of our knowledge. The extracted PPG features provide second level of classification between 0\% and 50\% force exertion levels in our work. For the cases where the face video signals predict the low levels (0\% or 50\%), the extracted features from the PPG signals are used to obtain efficient classification.


The proposed approach is shown to be robust to unseen data from an activity level that is different from the activities in our experiment. This is done by recording the face video and the PPG signals of the subjects at a new activity level corresponsing to when the subject is talking. The average movement landmarks for talking were extracted from the videos, and were used to predict force exertion level from our first model. If the first model predicts low, the extracted PPG features from the obtained PPG signal are passed through our second trained model. The results shows that first model classifies 7 out of 7 subjects belonging to low (0\% \& 50\%) force exertion level category and second model predicts 0\% force exertion level for 5 out of 7 subjects.

The rest of the paper is organized as follows. Section \ref{sec:setup} describes the experiment design for the data collection. Section \ref{sec:method} describes the methodology used in the paper to predict the force extraction levels. The two sets of neural networks are trained, one with features from the face videos and another with the features from the PPG signals. The combination of these two neural networks is used to classify the force exertion level into three categories. Section \ref{sec:results} presents the classification results of the two neural networks individually, as well as together. Further, the  model is shown to be robust as it is able to predict force exertion level well even even the participant is talking. Section \ref{sec:discussion} provides additional discussions for the obtained results. Section \ref{sec:conclusions} concludes the paper, with a brief mention of the potential future works.

\section{Experiment Design for Data Collection}\label{sec:setup}
\begin{figure}
	\centering
	\includegraphics[width=0.48\textwidth]{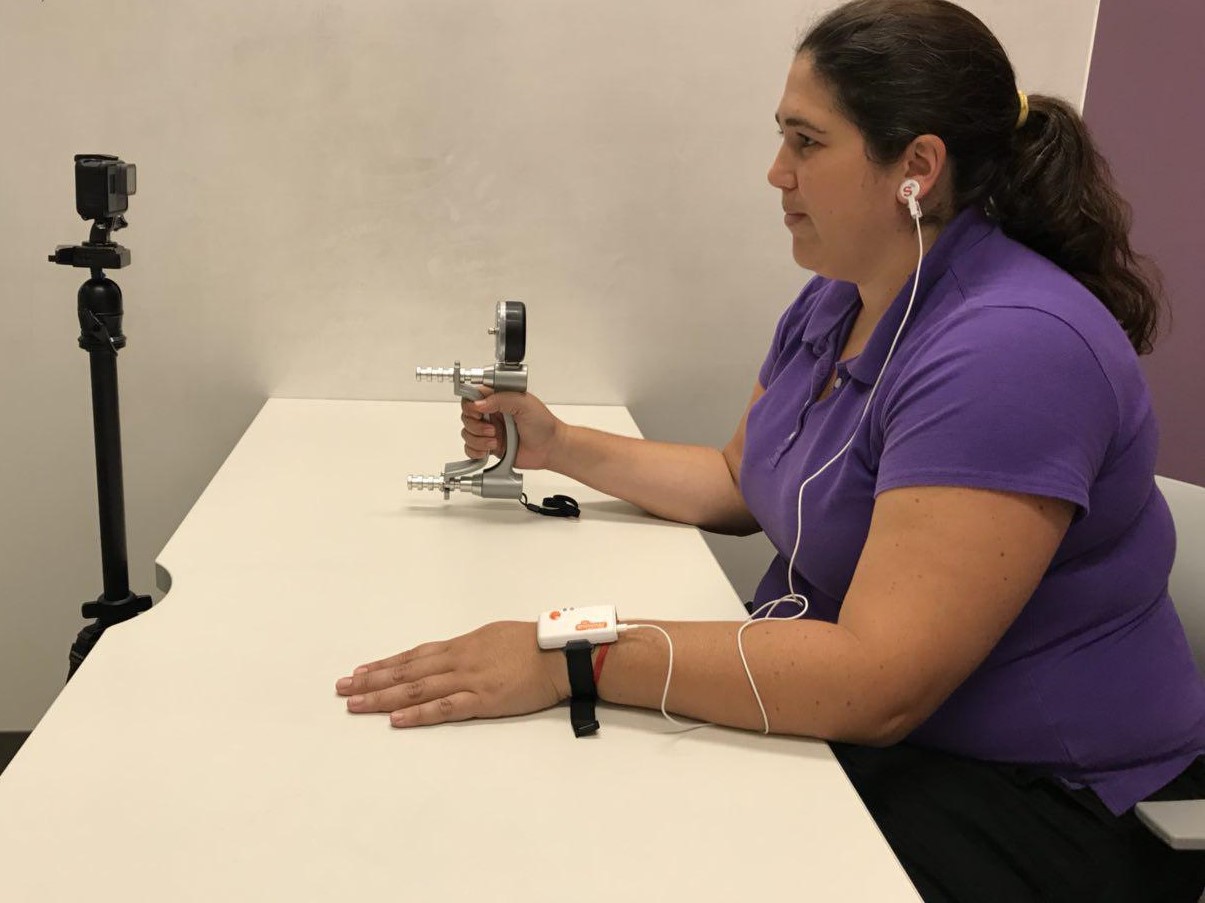}
	\caption{The experimental setup with a subject holding a grip dynamometer and pulse oximeter attached to the earlobe. The GoPro attached on the tripod is used to capture the video}
	\label{fig:exp}
\end{figure}

An experiment has been designed to collect the data that will be used to predict force exertion level. A study was conducted where each subject exerted varying levels of muscle force. During these activities, we  collected the videos of the person performing the activity as well as data regarding the volumetric blood flow by capturing PPG signal using pulse oximeter. Figure \ref{fig:exp} shows the complete set-up we used in our study.  

\subsection{Study participants}
Twenty healthy volunteers participated in this study. The participants were recruited from a university population through email including a description of the study. This study was reviewed by the university's Institutional Review Board and all participants provided informed consent. The only exclusion criteria were current injuries that prevented participants from performing force exertions. Sixteen males and 4 females participated in the study, all were right hand dominant, and their ages ranged from 18 to 29 years. The details of all the subjects that participated in the study is given in Table \ref{ref:tab1}. 

\subsection{Study Setup}
The power grip dynamometer was used to measure the  grip force of each subject. This devise helps in  measuring the maximum isometric strength of the hand and forearm muscles and hence helps us collecting the ground truth of force exertion level for each subject.


A GoPro camera was used to capture the video of our subjects while they were performing different kind of activities. We placed GoPro in front of the subject, around 0.5 meter away from face, and video recorded the subject during the entire experiment. It is a 12 MP camera and recordings are done at 50 frames per second.

The photoplethysmogram (PPG) signals were recorded using using pulse oximeter. The PPG signals were captured by Shimmer GST+. This device has a contact probe that is attached to the earlobe. The earlobe is chosen as the suitable position for recording the PPG.. Although the signals could be estimated using the non-contact methodologies \cite{kumar2015distanceppg}, in this study the actual PPG signals were recorded using pulse oximeter to minimize the errors of estimation. 

\begin{table}[htbp]

\centering
 \begin{tabular}[width=0.47\textwidth]{|c c c c|} 
 \hline
 & \multicolumn{3}{c|}{Female (n=4)}  \\
 \hline
 & Mean $\pm$ SD & Min & Max  \\ 
 \hline 
Age (years) & 20.0$\pm$1.4 & 19 & 22 \\
\hline
Weight (lb) & 124.0$\pm$33.9 & 100 & 148 \\
\hline
Grip Force (lb) & 47.0$\pm$15.4 & 30 & 62 \\
 \hline \hline
 
  &  \multicolumn{3}{c|}{Male (n=16)}  \\
\hline
 &  Mean $\pm$ SD & Min & Max \\
 \hline 
Age (years) &  20.8$\pm$2.7 & 18 & 29 \\ 
\hline
Weight (lb) &  133.8$\pm$21.7 & 110 & 168 \\

\hline
Grip MVC (lb) & 88.8$\pm$20.4 & 62 & 118 \\
 \hline
 \end{tabular}
\caption {Data for 20 subjects in our experiment}\label{ref:tab1}
\end{table}

\begin{figure*}
	\centering
	\includegraphics[width=0.7\textwidth]{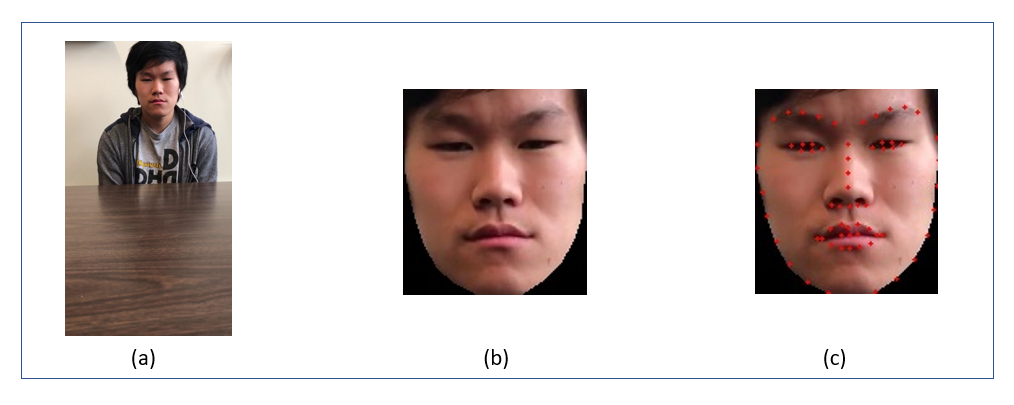}
	\caption{The steps followed for feature extraction from each frame of the video. (a) The actual image (one of the many frame) from the video captured during the experiment. (b) The detected and aligned face using DeepFace. (c) The face along with the 68 landmarks on it. These 68 landmark points are used by DeepFace in face recognition.}
	\label{fig:deep}
\end{figure*}

\subsection{Study Design}\label{ref:study}
At the beginning of the data collection session, participants were provided a description of the study, and written consent was collected. First, the subjects were seated in front of the white background to minimize the noise in video processing in detecting the face. The handheld dynamometer was calibrated as per the hand size to ensure standardized and comfortable gripping postures for each subject. This follows attaching pulse oximeter's contact probe  properly  to the subject's earlobe.

Participants were given a 5-minutes practice period to familiarize with the devices and environment. The overall study involved three different activity levels at different setting of grip dynamometer. In the first trial, each participant performed a grip exertion at maximum force that they are capable of. The subjects were instructed to maintain the maximum force for 9 seconds (note that although the magnitude of the force may decrease during the 9-seconds, participants continued to exert their maximum effort). The recordings were stopped after 9 seconds.The second exertion trial was 0\% grip force. In this trial, subjects were asked to hold the grip dynamometer without exerting any grip force.  The subjects rested for 5 minutes between each force exertion levels to prevent fatigue effects from carrying over to the next force exertion trial.  Finally, the last trial was force exertions at 50\% of maximum force . In this trial, each subject was asked to exert exactly 50\% of their maximum grip contraction. The distribution of the grip force for different subjects is reported in Table \ref{ref:tab1}.

\begin{figure}
	\centering\
	\includegraphics[width=0.48\textwidth]{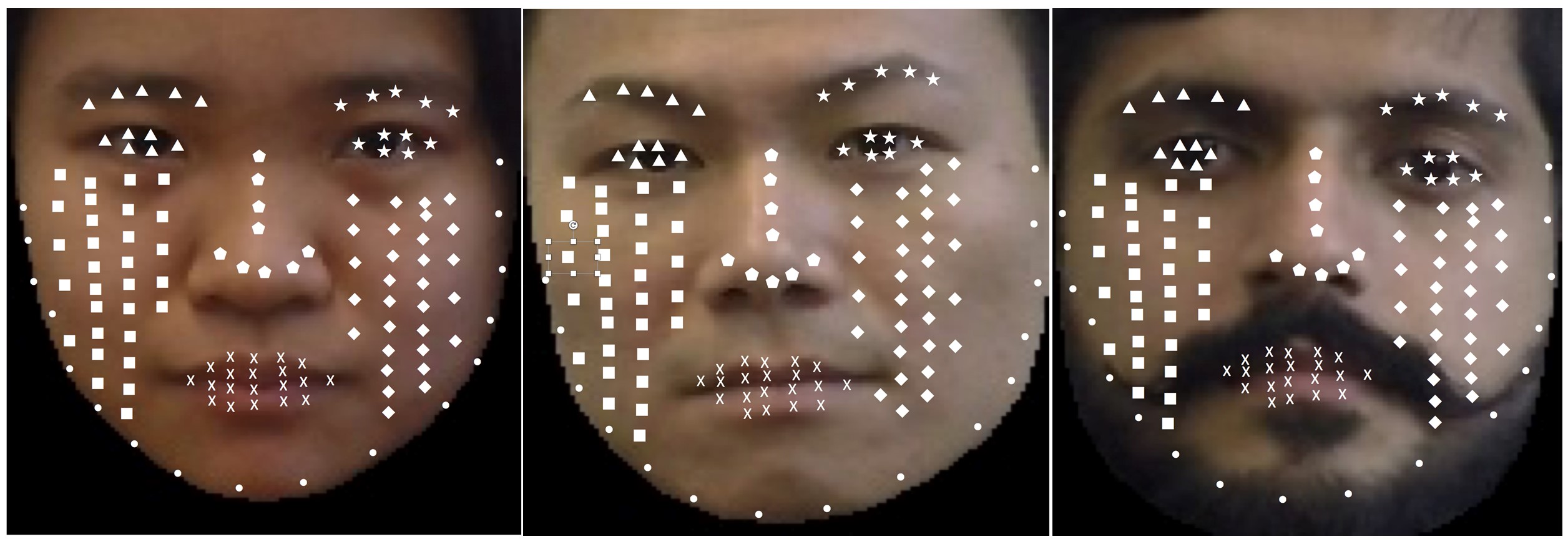}
	\caption{The location of 128 landmark points on the face for different subjects. Additional 60 landmarks have been identified on the face for efficient model training}
	\label{fig:128}
\end{figure}
\section{Methodology}\label{sec:method}

 The overall algorithm proposed in this paper takes person's video as an input and outputs the force exertion level of the person.  
 
 The methodology devised for predicting the force exertion level from human facial videos consists of using techniques of computer vision and machine learning. The overall method consists of 3 main steps: 
 First step involves processing the videos and extract meaningful and relevant parts of the video, i.e., cropped and aligned human face in our case. Second step extracts the important features from the different frames of the video and PPG signal followed by the third step in which we train neural network that will output a model to predict the force exertion level of person
 


The different steps  of the proposed method is discussed in details here.

\makeatletter
\def\BState{\State\hskip-\ALG@thistlm}
\makeatother


\subsection{Video Processing}\label{sec:video}
The videos of several subjects are recorded under different force exertion  levels as explained in section \ref{ref:study}. Each video is processed using the Deepface algorithm proposed in \cite{taigman2014deepface}. This is a state-of-the-art algorithm developed by researchers at Facebook. Deepface is a face recognition algorithm that consists of four main stages: 1. Detect 2. Align 3. Represent, and 4. Classify.

There have been other work in developing algorithm for  facial recognition \cite{barkan2013fast,cao2013practical,chen2013blessing,chopra2005learning,huang2012learning,sun2013deep}, but Deepface \cite{taigman2014deepface} reached an accuracy of 97.35\% in Labeled Faces in the Wild (LFW) dataset and reduced the error in face recognition of current state-of-the art by more than 27\%. The high accuracy in Deepface is achieved by  revisiting both alignment and representation step. 3D face alignment has been done using piecewise affine transformation and face representation is derived using 9-layer neural network which is a key for the high performance. Therefore, we utilized Deepface for recognizing faces in our approach. 

The 9 seconds video of each subject is trimmed to 7 seconds before passing it to Deepface. The first 2 seconds of videos are removed because each subject requires initial 1-2sec to reach to the required force level. Each video is recorded at 50 frames per second and hence, consists of 350 frames We process all these frames using Deepface that recognizes and aligns the face of each subject across the frames using 68 landmark points on the face. Figure \ref{fig:deep} shows how deepface is used to extract faces from the each frame in the video. Figure \ref{fig:deep} (a) is an example of an actual frame in the video. DeepFace recognizes the face of the person in each and crops the face out of it as shown in Figure \ref{fig:deep} (b).  This algorithm helps identify 68 landmark points on the face as depicted in Figure \ref{fig:deep} (c) and track these 68 landmark points over the whole video
The 68 landmark points represents the contour of the face, eyebrows, eyes, lips, and nose.  Detecting and aligning the face in each frame of the video 
is one of the most critical step in our overall methodology, because relevant features to train a neural network will be extracted from the output of Deepface. 



\begin{figure*}
	\centering
	\includegraphics[width=0.8\textwidth]{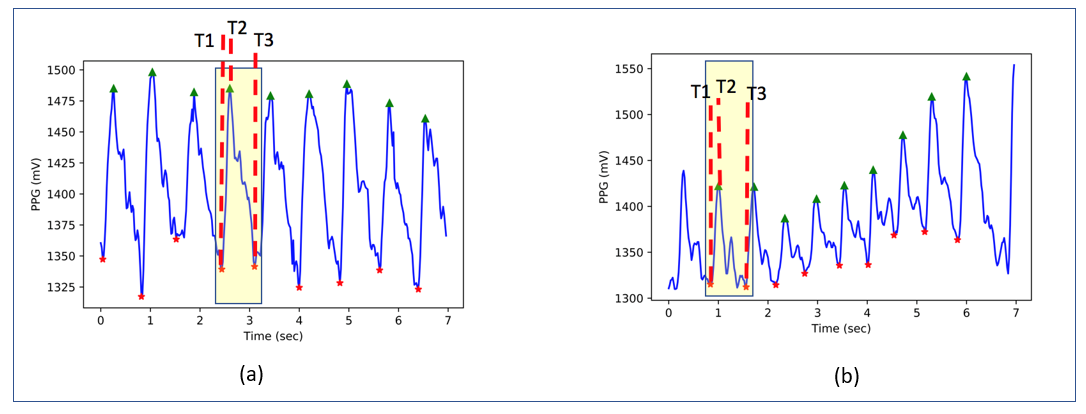}
	\caption{Collected PPG Signals (a) 0\% grip force (b)50\% grip force; T1: time at first local minimum of the beat, T2: time at local max of the beat, T3: time at the end of the beat}
	\label{fig:ppg}
\end{figure*}

\subsection{Feature Extraction}

The extraction  of ``right'' features is important as it plays significant role in training a neural network. The choice of relevant features leads to the simplification of the models which in turn requires shorter training time \cite{james2013introduction}. ``Right'' set of features helps in avoiding the curse of dimensionality and leads to generalization of the model by reducing the variance in the model \cite{bermingham2015application}. Choosing the subset of features from the available data reduces redundancy in the input to the neural networks and subsequently improving the performance.  We will extract relevant features from two sources: 1. Frames that has been processed by DeepFace 2. PPG signal that has been collected during our data collection process

\subsubsection{\textbf{Features from Videos}}\label{sec:landmarks}

Deepface utilizes the information of 68 landmark points on the face. Our proposed method use 128 landmark points on the face as shown in Figure \ref{fig:128}. Based on 68 landmark points, we locate 60 more landmarks on the face that lies on the left and right cheeks.  Thirty landmarks on each cheek is located based on the location of landmarks on the contour of the face and eyes. Different landmark points can be grouped together based on the location on the face as: 1: Contour of Face (17 landmarks), 2: Left Eye with left eye brow (11 landmarks), 3: Right eye with right eyebrow (11 landmarks), 4: Nose (9 landmarks), 5: Lips (16 landmarks), 6: Left Cheek (32 landmarks), 7: Right Cheek (32 landmarks).

All the 128 landmark points are tracked in 350 frames for each video. The location ($x$ and $y$ co-ordinate values) of each landmark is extracted and based on the location, the average movement of each landmark with respect to its location in the first frame  is calculated over the entire video. For each video, average movement, $d_j$,of each landmark, $j$, is given in  equation \ref{eq:feature}

\begin{equation}\label{eq:feature}
  d_j=   \dfrac{\sum_{i=1}^{n}\sqrt{(x_{ji}-x_{j1})^2 + (y_{ji}-y_{j1})^2}}{n}
\end{equation}

where $n$ is the number of frames in the video. Thus, the set $D_1$=\{$d_1,d_2.......d_{128}$\} is our first set of features used in the prediction of exertion level.
These features are potential indicator of exertion level as they depict how each point on the face moves in the entire video when subjects are asked to perform different exertion level activity.

\subsubsection{\textbf{Features from PPG}}\label{sec:ppg}
The other set of features is derived using PPG signal captured during our data collection experiments.
PPG signal of each subject consists of multiple beats where each beat is defined as the set of consecutive values of PPG having a maximum PPG value between two minimum values as highlighted in yellow color in Figure \ref{fig:ppg} (a). Therefore, for each beat we have a starting point denoted at T1, maximum point denoted as T2, and end point denoted at T3.
The total number of beats of 7 seconds recorded PPG were varied from 8 to 12 beats. 
Therefore, from each PPG signal following features are extracted:
\begin{itemize}
\item Time interval between T1 and T2 is extracted for first 5 beats
\item Time interval between T2 and T3 for  first 4 beats
\item Time interval between T1 and T3 for first 4 beats
\item Standard deviation of PPG values at T2 for all beats
\item Standard deviation of PPG values at T1 for all beats
\item Mean of three time intervals: T1 and T2, T2 and T3 and T1 and T3
\item Standard deviation of three time intervals: T1 and T2, T2 and T3 and T1 and T3
\end{itemize}

The above mentioned 21 features were extracted from PPG signal corresponding to each video. The set of these 21 features is referred to as $D_2$ in rest of the paper. The PPG features extracted from the signal  corresponds to the cardiovascular activities of the person during different experiment levels. In this work, we extracted these features from PPG signal that is captured using a contact device. Recently there have been advancements in passively estimating the PPG signal using facial video without the need of any contact device. There are many state-of-the-art techniques discussed in \cite{verkruysse2008remote,kumar2015distanceppg, poh2011advancements, poh2010non, sun2011motion, Wieringa2005, humphreys2007noncontact, holton2013signal} that utilizes videos of the human to extract PPG signal.

The features extracted for model training is our main novelty as none of the other authors have extracted such features and used machine learning to predict the force exertion level. This is the first ever work that utilizes such facial and PPG features.


\subsection{Model Training}
After all the relevant features $D_1$ and $D_2$ are extracted, we train two neural networks : $NN_1$ \& $NN_2$ to predict the three levels of exertion level of human. The feature set $D_1$ is used for training $NN_1$  that classifies 100\% force exertion level and rest of other levels (0\% \& 50\% ) and further feature set $D_2$ is trained on $NN_2$ to classify between 0\% and 50\% level.


The architecture of both $NN_1$ and $NN_2$ used to train the features is same as shown in Figure \ref{fig:arch}. The extracted features $D_1$ \& $D_2$ were used as the input data into a neural network with 1 input , 3 hidden and 1 output layers as shown in Figure \ref{fig:arch}. For each hidden layer, 35 neurons are used. The activation function used in the training of network is exponential linear units (\textit{ELUs}) \cite{DBLP:journals/corr/ClevertUH15} as defined in equation \ref{eq:elu}. Batch normalization is used in each hidden layer \cite{DBLP:journals/corr/IoffeS15}. The use of drop-out is one of the simplest way to avoid over-fitting of the neural network \cite{srivastava2014dropout}. Drop-out rate was set to 50\% to avoid over-fitting in all the three hidden layers. This will help in better generalizing the network for unseen data. In the output layer, two neurons were used for 100\% and rest (0\% and 50\%) for $NN_1$ \& 0\% force exertion and 50\% force exertion level for $NN_2$. The best performance of the network is achieved with using Adam \cite{DBLP:journals/corr/KingmaB14} as an optimizer along with categorical cross-entropy as a loss function.

\begin{equation}
f(x) = 
     \begin{cases}
       \text{x,} &\quad\text{if x}\ge0\\
       \alpha\times(e^x -1) &\quad\text{if x}\le0 \\
     \end{cases}
     \label{eq:elu}
\end{equation}




\tikzset{%
  every neuron/.style={
    circle,
    draw,
    minimum size=0.6cm
  },
  neuron missing/.style={
    draw=none, 
    scale=2,
    text height=0.2cm,
    execute at begin node=\color{black}$\vdots$
  },
}
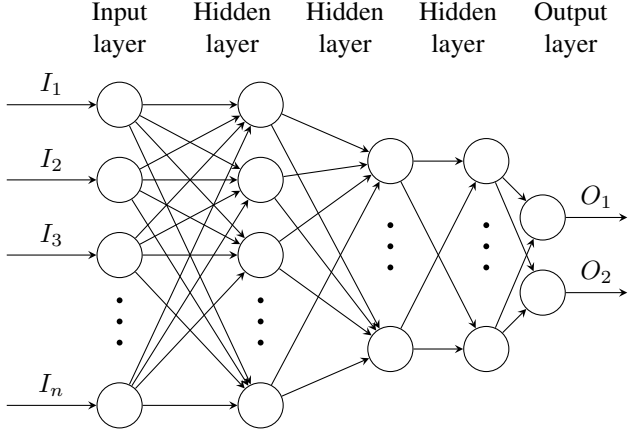
\begin{figure}
\centering\

\begin{tikzpicture}[x=1.5cm, y=1cm, >=stealth]

\foreach \m/\1 [count=\y] in {1,2,3,missing,4}
  \node [every neuron/.try, neuron \m/.try] (input-\m) at (0,2.5-\y) {};

\foreach \m [count=\y] in {1,2,3,missing,4}
  \node [every neuron/.try, neuron \m/.try ] (hidden1-\m) at (1.25,2.5-\y) {};

\foreach \m [count=\y] in {1,missing,2}
  \node [every neuron/.try, neuron \m/.try ] (hidden2-\m) at (2.4,2-\y*1.25) {};
  
\foreach \m [count=\y] in {1,missing,2}
  \node [every neuron/.try, neuron \m/.try ] (hidden3-\m) at (3.25,2-\y*1.25) {};

\foreach \m [count=\y] in {1,2}
  \node [every neuron/.try, neuron \m/.try ] (output-\m) at (3.75,1-\y) {};

\foreach \l [count=\i] in {1,2,3,n}
  \draw [<-] (input-\i) -- ++(-1,0)
    node [above, midway] {$I_\l$};


\foreach \l [count=\i] in {1,2}
  \draw [->] (output-\i) -- ++(0.75,0)
    node [above, midway] {$O_\l$};

\foreach \i in {1,...,4}
  \foreach \j in {1,...,4}
    \draw [->] (input-\i) -- (hidden1-\j);
    
\foreach \i in {1,...,4}
  \foreach \j in {1,...,2}
    \draw [->] (hidden1-\i) -- (hidden2-\j);

\foreach \i in {1,...,2}
  \foreach \j in {1,...,2}
    \draw [->] (hidden2-\i) -- (hidden3-\j);

\foreach \i in {1,...,2}
  \foreach \j in {1,...,2}
    \draw [->] (hidden3-\i) -- (output-\j);

\foreach \l [count=\x from 0] in {Input, Hidden, Hidden, Hidden, Output}
  \node [align=center, above] at (\x,2) {\l \\ layer};

\end{tikzpicture}  
  \caption{The architecture of a fully connected neural network with three hidden layers} \label{fig:arch}

\end{figure}  

\begin{table*}
\centering
\begin{adjustbox}{width=\textwidth}
\begin{tabular}{ |c|c|c|c|c|c|c|c|c|c|c|c|c|c|c|c|c|c|c|c|c| } 
 \hline
 Subject ID & 1&2&3&4&5&6&7&8&9&10& 11&12&13&14&15&16&17&18&19&20  \\ 
Group A & o &o &o &o &o &o &o &o &o &o &o &o &x &x &x &o &o &o &o &o    \\ 
Group B &  o &o &o &o &o &x &o &o &o &o &o &o &o &o &o &o &o &o &o &o    \\ 
 
 \hline
\end{tabular}
\end{adjustbox}
\caption{Table showing prediction results for $NN_1$}\label{tab:NN1}
\end{table*}
\section{Results}\label{sec:results}

This section summarizes the classification results of both neural networks i.e., $NN_1$ \& $NN_2$.  The performance of final model is discussed  along with the noise analysis that shows the robustness of the trained model. 

\subsection{Force level classification using \textbf{$D_1$}}
\begin{figure}[htbp]
	\centering\
	\includegraphics[width=0.48\textwidth]{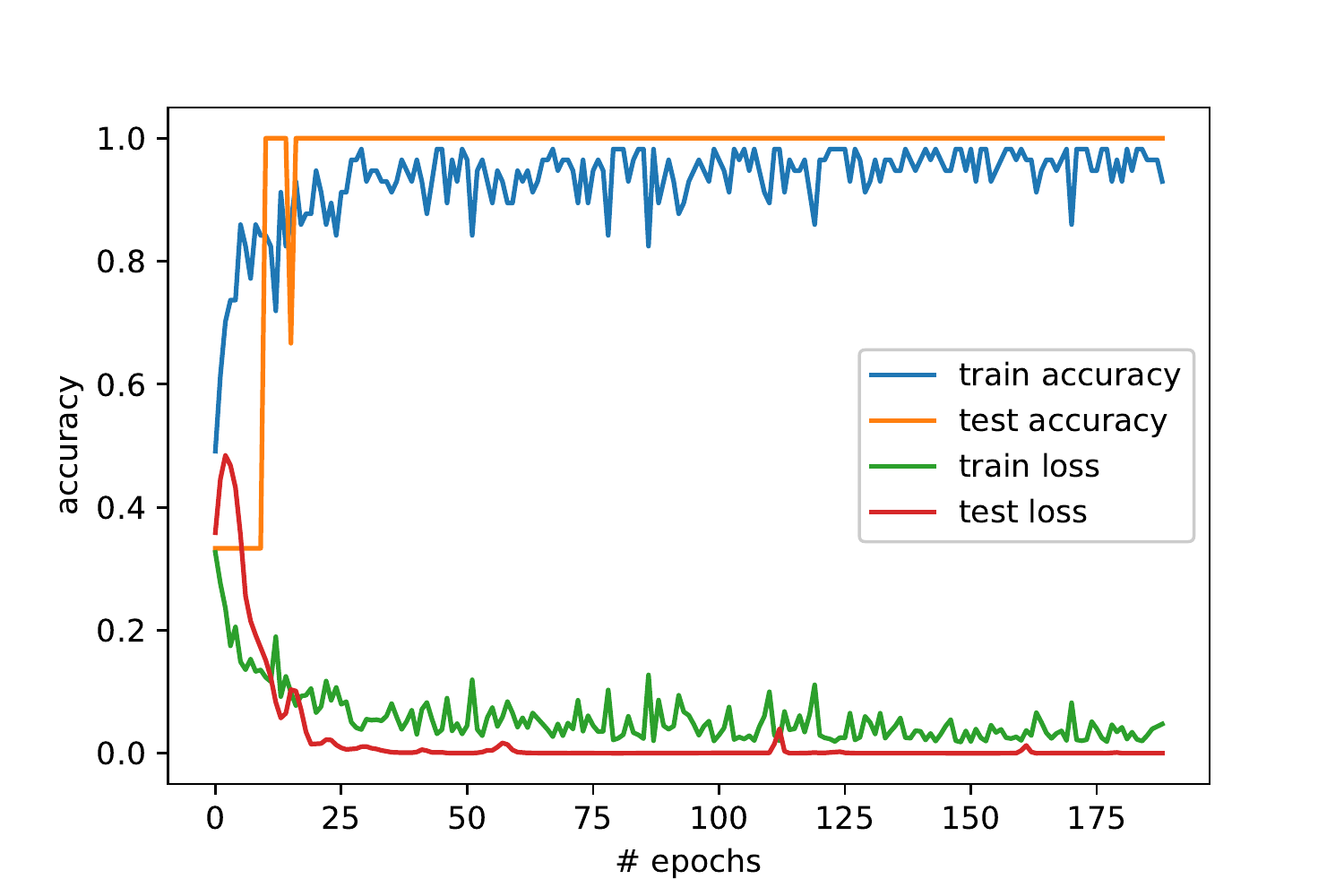}
	\caption{The behavior of accuracy and loss values of $NN_1$ for test and train dataset against number of epochs}
	\label{fig:NN1}
\end{figure}

The classification between 100\% (group A) and 0\% \& 50\% (group B) is done using $NN_1$. The model is trained using a fully connected neural network as architecture that utilizes average movement of landmark points. The neural network is trained for 200 epochs. The number of epochs are decided based on the performance of test and train loss curves . The test loss for the performance of neural network is reported using leave one out cross validation approach meaning during training the neural network, we use data for average movements of all force level for 19 subjects and once the model is trained, the performance is measured using data from 1 subject that has been left out of training. This approach has been repeated for all the subjects and the average accuracy on test set has found to be 90\% . The graph in figure \ref{fig:NN1} represents the behavior of accuracy and loss value for test and training dataset for one of the subject. Table \ref{tab:NN1} shows our accuracy results for all subjects. The subjects with 'o' represents that they have been correctly classified for a particular group and subjects having 'x' corresponding to a particular group shows that the class is predicted incorrectly. We can predict Group B correctly for all the subjects except subject no. 6 and also we predict group A correctly 17 out of 20 subjects leading to an overall accuracy of 90\%.

\subsection{Force level classification using \textbf{$D_2$}}
\begin{figure}[htbp]
	\centering\
	\includegraphics[width=0.48\textwidth]{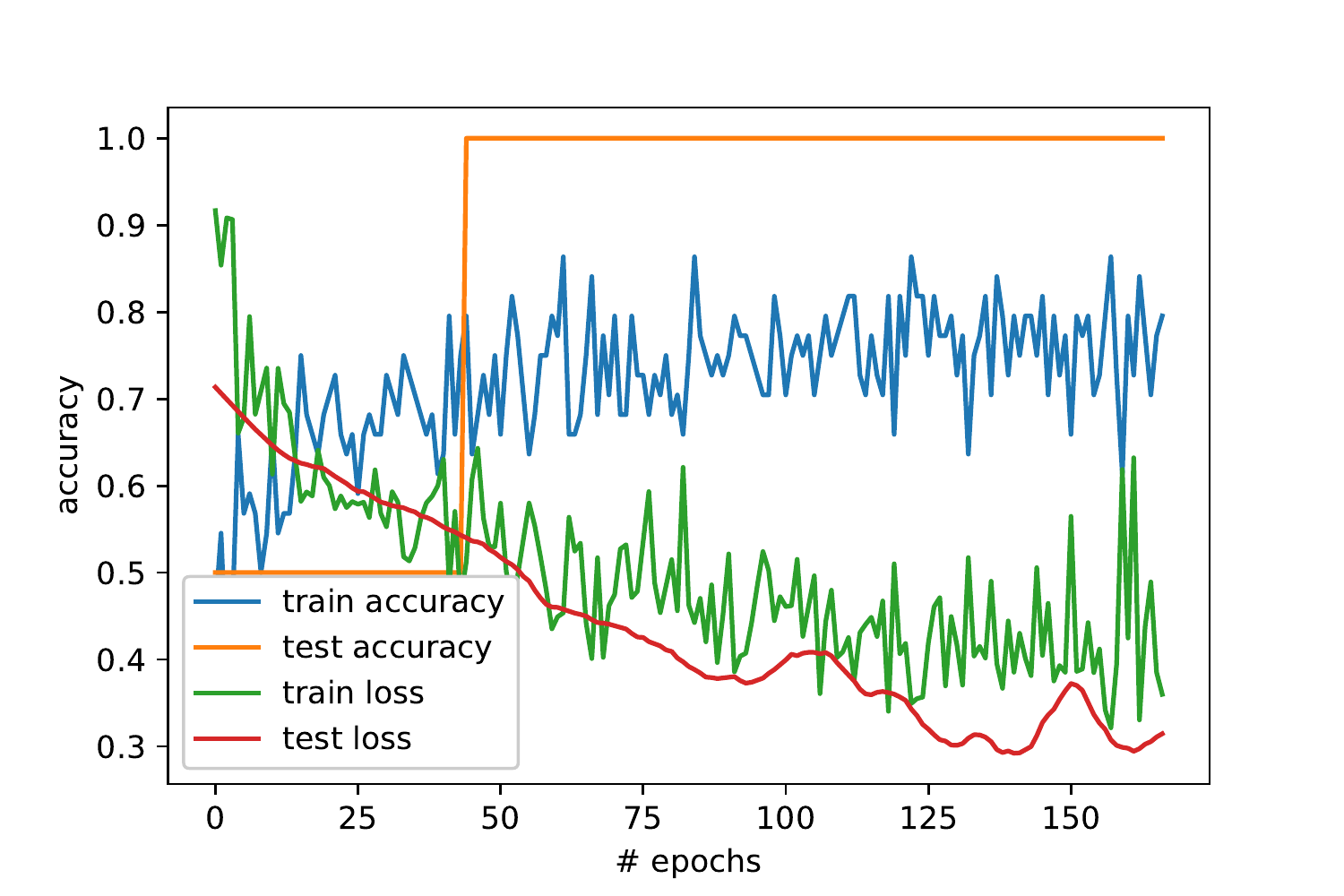}
	\caption{The behavior of accuracy and loss values of $NN_2$ for test and train dataset against number of epochs}
	\label{fig:NN2}
\end{figure}

After classifying Group A and Group B using $NN_1$, we use $NN_2$ to classify the 0\% and 50\%  force exertion level in Group B. This neural network model utilized all the features extracted from PPG that has been described in section \ref{sec:ppg}. This neural network is trained for 175 epochs. The number of epochs are chosen such that model doesnot ovefit. The technique of early stopping \cite{caruana2001overfitting} is used here to avoid over-fitting, reduce variance in the model and generalize model well over unseen data. This model also uses same approach of "leave one out" as has been discussed in previous subsection. The average accuracy on 20 subjects is 80\% for $NN_2$. The behavior of the accuracy and loss values for testing and training data while training the neural network is shown in Figure \ref{fig:NN2}. 
Table \ref{tab:NN2} shows our average accuracy results for each subjects. The subjects with 'o' represents that they have been correctly classified for a particular group and subjects having 'x' corresponding to a particular group shows that the class is predicted incorrectly. The model can correctly predict 0\% force exertion level for 18 out of 20 subjects and 50\% force exertion level for 14 subjects out of total of 20 leading to an overall accuracy of 80\%.

\begin{table}
\begin{adjustbox}{width=0.47\textwidth}
\begin{tabular}{ |c|c|c|c|c|c|c|c|c|c|c|c|c|c|c|c|c|c|c|c|c| } 
 \hline
 Subject ID & 1&2&3&4&5&6&7&8&9&10 \\ 
Group B-0\% & o &o &o &x &o &x &o &o &o &o   \\ 
Group B-100\% &  o &o &x &o &o &o &x &o &x &o   \\ 
 \hline \hline
 
  Subject ID &11&12&13&14&15&16&17&18&19&20 \\ 
Group B-0\%  &o &o &o &o &o &o &o &o &o &o   \\ 
Group B-100\% &o &o &x &x &o &o &o &x &o &o   \\ 
 \hline
\end{tabular}
\end{adjustbox}
\caption{Table showing prediction results for $NN_2$}\label{tab:NN2}
\end{table}

\subsection{Final Model}
Our methodology breaks the process of classifying different force exertion levels into two steps and we have provided our accuracy results for two scenarios separately. The overall accuracy of the two models combined is calculated to be 81.7\%. The predictions for all the subjects has been combined together from $NN_1$ and $NN_2$ and has been shown in Table \ref{tab:final}. 
There were some subjects that had same facial expression during both 50\% and 0\% force exertion level experiment and there was no significant difference between average movement of the 128 landmarks between these two levels
Hence, it is not advisable to use average movement of the landmark points as classifying feature between 0\% \& 50\% force exertion level. Therefore,  we extracted PPG features that relates the cardiovascular parameters with force exertion level.

\subsection{Model Robustness}

In order to check the robustness of our model, we collected the data of 7 additional subjects while they were performing different kind of activity. We captured the videos of the subjects using the same experimental set-up as explained in section \ref{sec:setup} when they were not performing any sort of force exertion activity, but are talking. We made each subject to speak a paragraph on themselves for 9 seconds for and we recorded their video along with PPG data during this activity. We name this activity as "talking". 

Using the same processing technique, we extracted $D_1$ and $D_2$ set of features for all the subjects during talking. The set $D_1$ is passed through trained $NN_1$ and set $D_2$ is passed through trained $NN_2$  and predictions are made as of what activity level they belong to. 

It is interesting to note that for all the 7 subjects when $D_1$ is passed through $NN_1$, it always predicts group B for the activity level which means that our algorithm is able to differentiate between talking and 100\% and therefore gives high probability to group B. When set of features derived from PPG are used as an input to the trained $NN_2$, for 5 out of 7 subjects, the model predicted it to be 0\% force exertion level and for 2 out of 7  subjects, model predicts as if subjects are at their 50\% force exertion level. Note that the data corresponding to talking was not used while training the network. It is completely unseen data for our two trained neural networks.

\section{Discussion}\label{sec:discussion}

We demonstrate that the techniques of computer vision and machine learning can predict the force exertion level using extracted features and provides a novel approach for such estimation. Understanding force exertion levels has important implications across domains and applications, and in this work, we demonstrate the approach in the context of workplace injuries. Specifically, varying levels of force and duration/frequency of these forces are predictive of musculoskeletal injuries. This section provides more discussion on using machine learning in prediction of force exertion level and provides more insights on the feature selection that we performed in our work.

\begin{table}
\begin{adjustbox}{width=0.47\textwidth}
\begin{tabular}{ |c|c|c|c|c|c|c|c|c|c|c|c|c|c|c|c|c|c|c|c|c| }
 \hline
Subject ID & 1&2&3&4&5&6&7&8&9&10 \\ 
0\% grip Force  & o &o &o &o &o &x &o &o &o &o   \\ 
50\% grip Force &  o &o &x &x &o &o &x &o &x &o   \\
100\% grip Force & o &o &o &o &o &o &o &o &o &o   \\ 
 \hline \hline
 
 Subject ID &11&12&13&14&15&16&17&18&19&20 \\ 
0\% grip Force &o &o &o &o &o &o &o &o &o &o  \\ 
50\% grip Force &o &o &x &x &o &o &o &x &o &o  \\
100\% grip Force &o &o &x &x &o &o &o &x &o &o  \\ 
 \hline
\end{tabular}
\end{adjustbox}
\caption{Table showing final prediction results from $NN_1$ \& $NN_2$}\label{tab:final}
\end{table}
\subsection{Machine Learning in Classifying Force Levels}
 The use of machine learning is two-fold in this work. First, the machine learning is used in DeepFace algorithm for facial recognition that our team further augmented with increased number of features. Secondly, we use machine learning to generate a classifier to predict different force exertion levels.

There are various methodologies \cite{5771424,cao2013practical,chen2013blessing,barkan2013fast,sun2013deep,chopra2005learning,huang2012learning,M.Osadchy2007Sfda} proposed that can achieve facial recognition but the methodology proposed in \cite{taigman2014deepface} outperforms other methods and results in the accuracy of 97.35\% in Labeled Faces in the Wild (LFW) dataset, reducing the error in face recognition of current state-of-the-art by more than 27\%. This method is more robust and the explanation on DeepFace is  discussed in section \ref{sec:video}. The 9 layer neural network used in Deepface makes it more robust to detect faces in the video for our study and henceforth extract relevant features from the video frames. These facial features represent a key component for force classification. 

The second use of machine learning is force classification. In this step, we added additional novelty by leveraging the underlying physiological mechanisms of generating muscle forces to improve force classification accuracy. Thus, we included features from PPG as well and deployed a fully connected neural network to train a model that can distinguish between different force exertion levels. The neural networks are known to be universal approximators \cite{hornik1989multilayer} and hence we use them to identify the underlying function explaining the relationship between the features and response variable.

\begin{figure}[htbp]
	\centering
	\includegraphics[width=0.48\textwidth]{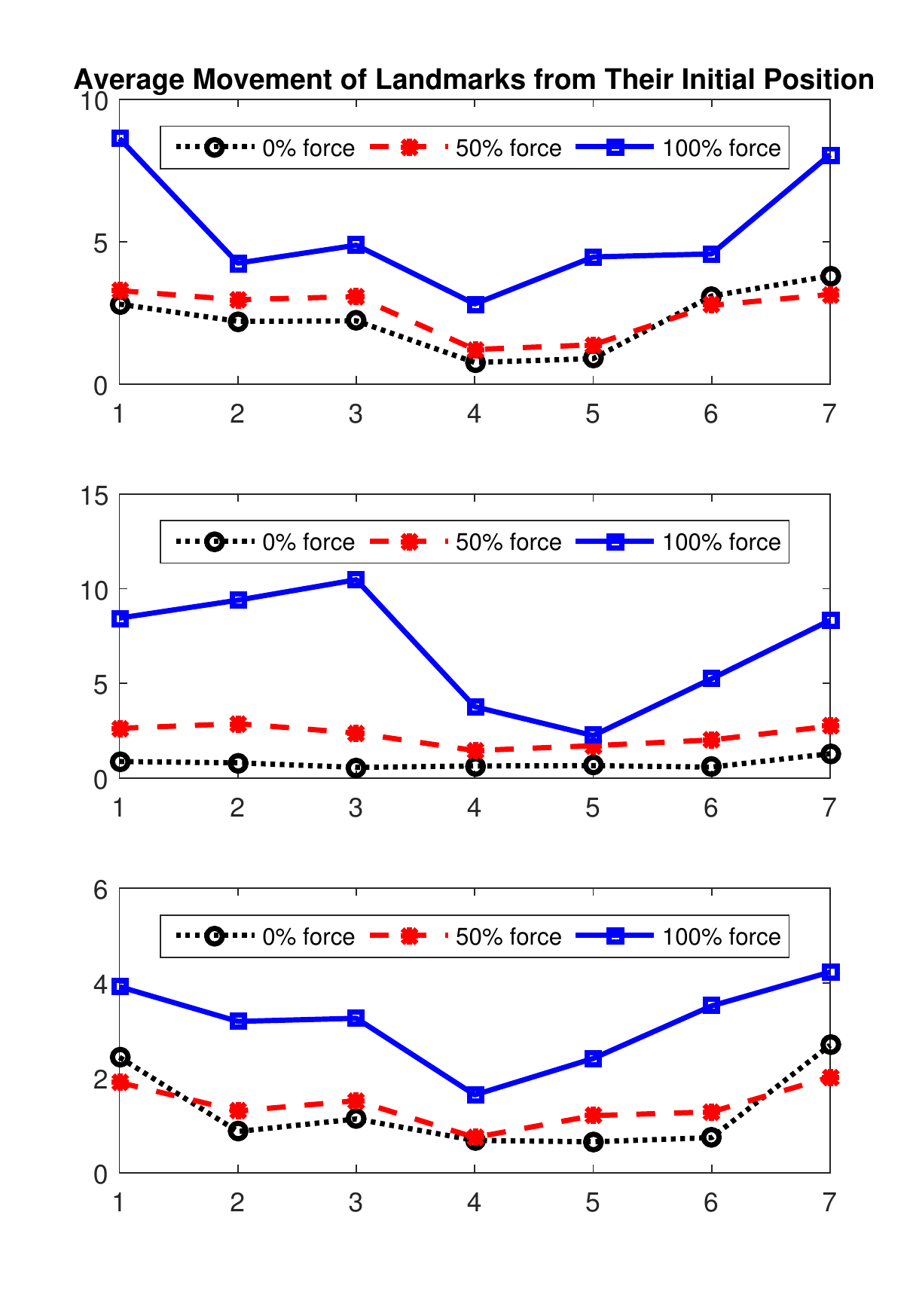}
	\vspace{-.5in}
	\caption{Variation in facial features groups of three randomly chosen subjects, 1: Contour of a face, 2: Left Eye + Eyebrow, 3: Right Eye + Eyebrow, 4: Nose, 5: Lips,6: Left Cheek, 7: Right Cheek}
	\label{fig:Capture}
\end{figure}

\begin{figure}[htbp]
	\centering
		\vspace{-.2in}
	\includegraphics[width=0.48\textwidth]{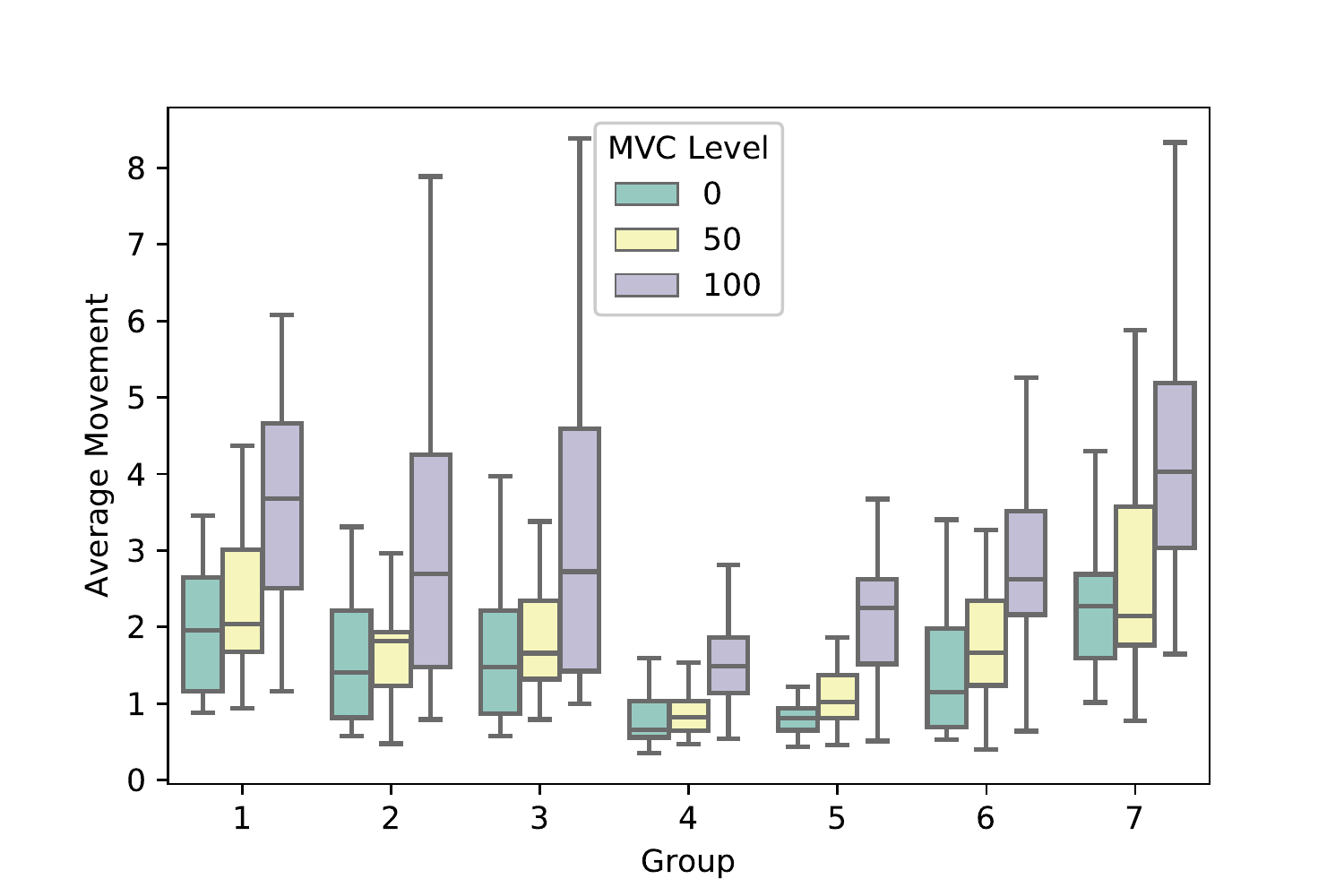}
		\vspace{-.2in}
	\caption{Variation in facial features groups of all subjects, 1: Contour of a face, 2: Left Eye + Eyebrow, 3: Right Eye + Eyebrow, 4: Nose, 5: Lips,6: Left Cheek, 7: Right Cheek}
	\label{fig:boxplot}
\end{figure}

\begin{figure}[htbp]
	\centering\
	\includegraphics[width=0.48\textwidth]{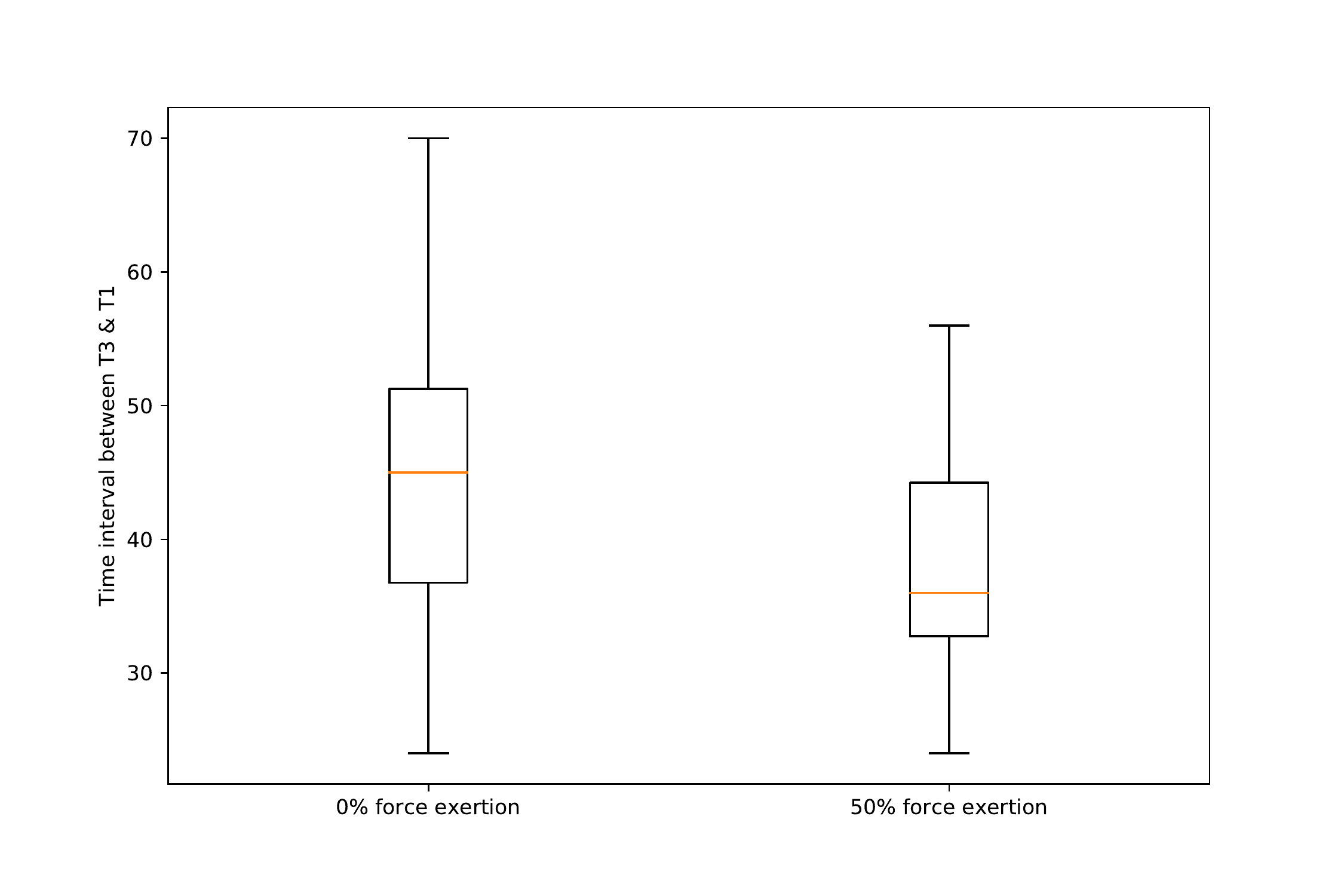}
	\caption{Variation in the time interval between T3 \& T1 for all subjects corresponding to 0\% \& 50\% force exertion level }
	\label{fig:ppg_t3}
\end{figure}

\subsection{Facial Features Selection}
The average movement of detected facial landmarks along with the cardiovascular features derived from PPG in different exertion levels have been used to classify the force exertion levels. Our novelty lies in choosing these relevant facial features. As the person increases her effort level, facial expression tends to change and there are differences in the average movement of the facial landmarks for different force exertion level. The identification of these visual cues were drawn for tools and techniques from the field of human factors. Specifically, ergonomics practitioners are trained to associate (through observation) cues like "Substantial Effort with Changed Facial Expression" with an MVC of 70 \% and very strong effort. In contrast, "Obvious Effort, But Unchanged Facial Expression" is associated with 40 \% MVC and moderate effort \cite{armstrong2007acgih}

Figure \ref{fig:boxplot} \& Figure \ref{fig:Capture} shows how different groups of facial landmarks behave differently for three different force exertion levels. Figure \ref{fig:Capture} shows the average movement of landmark groups for three randomly picked subjects. It is interesting to note that landmarks belonging to nose always shows least movement in all the three force exertion levels. On the other hand, face contour, eyes and cheeks show high average movements over the entire video. Figure \ref{fig:boxplot} generalizes this behavior over all the subjects and depicts the box plot of each force exertion level for all the 7 groups of landmarks. The change in the location of the landmarks on the face is explained by the motion of the muscles beneath the skin of the face. As body changes its actions, it leads to the changes in the facial expression \cite{FantoniCarlo2014BACt} and thus we observe the movement of landmarks for different force exertion levels. It is further interesting to note that average movement of landmarks are robust against day to day variations like change in the lighting around them, presence of make-up on the face etc. as well as robust for different people belonging to different skin tone. Therefore, choice of such facial feature leads to high accuracy of our model and make it robust for classifying higher force exertion level. 

As noted previously, changes in facial expressions are observed typically for strong exertions (~60-70 \% MVC); however, a known musculoskeletal injury mechanism is continuous and prolonged sub-maximal force exertions. Specifically, although a single moderate (~30-50 \% MVC) exertion may not lead to immediate injuries, repeated and prolonged exertions at these levels lead to cumulative trauma disorders. The key challenge is that facial expressions are more likely to be unchanged during these exertion levels. Thus, we further distinguished lower levels of force exertions using cardiovascular parameters of the person  which would be captured from PPG signal. This changing trend can be seen in Figure \ref{fig:ppg} where we observe the increasing trend of PPG signal for 50\% force exertion level and a stationary signal for  0\% force exertion level. 
Figure \ref{fig:ppg_t3} shows the observed variation in one of the PPG feature i.e., time interval between T3 and T1 for first four beats, for 0\% and 50\% force exertion levels. The mean and standard deviation of time interval between T3 and T1 for 0\% effort level is 1.03s and 0.55s respectively where mean and standard deviation for 50\% force exertion level is 0.86 sec and 0.52 sec respectively. Higher force exertion activity increases the heart rate of the person because of faster cardiac cycles, hence we see differences in PPG extracted features between 0\% \& 50\% effort level. Therefore, both average movement and PPG features become important features in our study. 

For our analysis, we utilized cardiovascular features derived from the PPG signal that had been captured using a contact device placed on the earlobe of the subjects. Although this technique requires contact, continued innovation in wearables (e.g., fitness watches and activity trackers) has provided many options for collection continuous PPG signals without significant cost to employers or usability/workflow burden to workers. Furthermore, over the last decade, there have been ongoing research in developing methods for estimating PPG signals from the facial videos using non contact methodology. The authors in \cite{humphreys2007noncontact} provided a technique that extracts PPG signal from the human facial videos using complimentary metal-oxide semiconductor camera with the use of external light emitting diodes. Also, the authors of  \cite{verkruysse2008remote, poh2011advancements} demonstrated that PPG signal can be estimated by just using ambient light as a source of illumination along with simple digital camera. Further advancements led to the formulation of more robust methodology that overcomes challenges in extracting PPG for people having dark skin tones \cite{kumar2015distanceppg}. There are many existing methods that can be easily used to derive PPG signal directly from the facial videos. Future work incorporating these techniques have the potential to make our proposed methodology completely passive and non-contact.





\subsection{Non-contact Exposure Assessment }

The force exertions has been considered as one of the main contributing factors in current risk assessment tools \cite{hignett2004rapid, mcatamney1993rula, steven1995strain}. The high variability of the identified risk score with respect to the estimated force exertion parameters is reported in current assessment tools. For example, the Strain Index Assessment \cite{steven1995strain} score will double if the intensity of the exertion changes from 20\% to 40\% \cite{Koppelaar2005}. In addition, Bao et al. reported weak correlation values between the ergonomists estimates and the worker's self-reports for pinch and grip force. Further exploration suggested among relationships of worker's self-reports, the ergonomist's estimates and the directly measured hand forces \cite{bao2006quantifying}. The proposed non-contact assessment method for classifying force levels can provide an objective automated estimations of hand forces.

\section{Conclusions}\label{sec:conclusions}
We demonstrate that a computer vision approach is effective in detecting force exertions across individuals. The approach proposed in this work is robust and model can be used for any new subject for such predictions. Computer vision based monitoring have shown effective applications in fall detection and health monitoring, and this work presents a computer vision framework for musculoskeletal injuries. Because computer vision is not intrusive to the workers and can be done without the need for specialize equipment, this technique will provide workplaces a transformative tool for ensuring on-the-job force requirements (effort level and duration at these levels) do not contribute to workplace injuries.  Although current work accurately classifies three force exertion levels, work is ongoing to expand this technique to other exertion levels to better meet the varying needs of different workplaces.

\section{Acknowledgments}

The authors would like to thank Lingjun Chen and Chufan Gao at Purdue University for the help in data collection and running experiments.  We gratefully acknowledge the support of NVIDIA Corporation with the donation of the Titan Xp GPU used for this research.

\bibliographystyle{IEEEtran}

\bibliography{bibfile}
\end{document}